\documentclass[prd, reprint, preprintnumbers,
twocolumn,
eqsecnum,floatfix,letterpaper,superscriptaddress,nofootinbib]{revtex4}
\usepackage{mathtools}
\usepackage{amsmath,amssymb,graphicx}
\usepackage[printwatermark]{xwatermark}
\usepackage{xcolor}
\usepackage{lipsum}
\usepackage{bm} 
\usepackage{color}
\usepackage{booktabs}
\usepackage{tabularx}
\usepackage{times}
\usepackage{microtype}
\usepackage{booktabs}
\usepackage{subfigure}
\usepackage{csquotes}
\usepackage{graphicx}
\usepackage{array}
\newcolumntype{M}[1]{>{\centering\arraybackslash}m{#1}}
\newcolumntype{N}{@{}m{0pt}@{}}

\graphicspath{ {/Users/sayantanidatta/Documents/projects/tgr_svd}}
\usepackage[normalem]{ulem}
\usepackage[varg]{txfonts}
\usepackage{pdfpages}
\usepackage{dirtytalk}
\usepackage[colorlinks, pdfborder={0 0 0}]{hyperref}
\definecolor{LinkColor}{rgb}{0.75 , 0, 0}
\definecolor{CiteColor}{rgb}{0, 0.5, 0.5}
\definecolor{UrlColor}{rgb}{0, 0, 0.75}
\hypersetup{linkcolor=LinkColor}
\hypersetup{citecolor=CiteColor}
\hypersetup{urlcolor=UrlColor}

\allowdisplaybreaks
\newcommand{\mycomment}[1]{}

\begin{document}
\title{Multiparameter tests of general relativity using principal component analysis with next-generation gravitational wave detectors}
\author{Sayantani Datta}\email{sdatta94@cmi.ac.in} 
\affiliation{Chennai Mathematical Institute, Siruseri, 603103, India}
\author{M. Saleem}\email{mcholayi@umn.edu} 
\affiliation{School of Physics and Astronomy, University of Minnesota, Minneapolis, MN 55455, USA}
\author{K. G. Arun}\email{kgarun@cmi.ac.in} 
\affiliation{Chennai Mathematical Institute, Siruseri, 603103, India}
\author{B. S. Sathyaprakash}  \email{bss25@psu.edu} 
\affiliation{Institute for Gravitation and the Cosmos, Department of Physics, Penn State University, University Park, Pennsylvania 16802, USA}
\affiliation{Department of Astronomy and Astrophysics, Penn State University, University Park, Pennsylvania 16802, USA}
\affiliation{School of Physics and Astronomy, Cardiff University, Cardiff, CF24 3AA, United Kingdom}
\date{\today}
%%%%%%%%%%%%%%%%%%%%%%%%%%%%%%%%%%%%%%%%%%%%%%%%%%%%%%%%%%%%%%%%%%%%%%%%%%%%%%%%%%%%%%%%%%%%%%%%%%%%%%%%%%%%%%%%%%%%%%%%%%%%%%%%%%%%%%%%%%%%%%%%%%%%%
\begin{abstract}
Principal Component Analysis (PCA) is an efficient tool to optimize the multiparameter tests of general relativity (GR) where one tests for simultaneous deviations in multiple post-Newtonian (PN) phasing coefficients by introducing fractional deformation parameters. We use PCA to construct the `best-measured’ linear combinations of the PN deformation parameters from the data. This helps to set stringent limits on deviations from GR and detect possible beyond-GR physics. In this paper, we study the effectiveness of this method with the proposed next-generation gravitational wave detectors, Cosmic Explorer (CE) and Einstein Telescope (ET). Observation of compact binaries with total masses between 20--200 $\mathrm{M}_{\odot}$ in the detector frame and at a luminosity distance of 500 Mpc, CE can measure the three most dominant linear combinations to an accuracy better than 10\%, and the most dominant one to better than 0.1\%. For specific ranges of masses and linear combinations, constraints from ET are better by a few factors than CE. This improvement is because of the improved low frequency sensitivity of ET compared to CE (between 1--5 Hz). In addition, we explain the sensitivity of the PCA parameters to the different PN deformation parameters and discuss their variation with total mass. We also discuss a criterion for quantifying the number of most dominant linear combinations that capture the information in the signal up to a threshold.
\end{abstract}
	
 \maketitle
%%%%%%%%%%%%%%%%%%%%%%%%%%%%%%%%%%%%%%%%%%%%%%%%%%%%%%%%%%%%%%%%%%%%%%%%%%%%%%%%%
%%%%%%%%%%%%%%%%%%%%%%%%%%%%%%%%%%%%%%%%%%%%%%%%%%%%%%%%%%%%%%%%%%%%%%%%%%%%%%%%%
\section{Introduction}\label{intro}
%%%%%%%%%%%%%%%%%%%%%%%%%%%%%%%%%%%%%%%%%%%%%%%%%%%%%%%%%%%%%%%%%%%%%%%%%%%%%%%%%
Advanced LIGO~\cite{aLIGO} and Advanced Virgo~\cite{aVirgo} have observed many gravitational wave (GW) signals from compact binary coalescences~\cite{Discovery,GW151226,GW170104,GW170608,GW170814,GWTC1,Venumadhav:2019lyq,Zackay:2019btq,Abbott:2020niy} during their first three observing runs, providing a unique opportunity to test the validity of Einstein’s theory of general relativity (GR) in the strong field regime~\cite{WillLR05,YunesSiemens2013,GairLivRev,Berti:2015itd}. To date, the gravitational wave observations have shown an excellent agreement with the predictions of GR~\cite{O1BBH,O2-TGR,Abbott:2020jks}. However, the planned upgrades in the sensitivities of Advanced LIGO and Virgo~\cite{OSD}, along with prospects of KAGRA~\cite{KAGRA} and LIGO-India~\cite{Saleem:2021iwi} joining the network of GW detectors, will allow us to carry out tests of GR with greater precision. 

One of the theory-agnostic tests of GR that is routinely performed on GW signals is the {\it parametrized tests of GR}~\cite{AIQS06a, AIQS06b, MAIS10, YunesPretorius09, LiEtal2011, TIGER, YYP2016}. This is a set of null tests in which the phase of the gravitational wave signal, as predicted by the Post-Newtonian (PN) approximation to GR, is deformed at every order by introducing fractional deformation parameters. These deformation parameters capture potential deviations from GR via the physical effects that appear at different PN orders~\cite{Bliving} and are absent in GR. This parametrization assumes that a modified theory of gravity will result in phase evolution of the emitted signal different from that in GR~\cite{AlexanderYunes07a,AlexanderYunes07b,CSreview09}. Such deviations from GR at various PN orders may be detectable by measuring these phenomenological PN deformation parameters and testing for their consistency with the predictions of GR~\cite{AIQS06a,AIQS06b,Chamberlain:2017fjl,Barausse:2016eii}.

In this paper, we will mainly look at tests of GR using the inspiral phase of the gravitational wave signal. During this phase, the binary components are orbiting towards each other under gravitational radiation reaction, adiabatically. Higher order PN coefficients in the formula for the phase evolution contain multiple nonlinear interactions (such as tails~\cite{BD88,BS93}, tails of tails~\cite{B98quad,B98tail} etc) and physical effects (such as spins-orbit and spin-spin interactions~\cite{KWWi93,K95,BBuF06,BFH2012}). Further, most of these effects appear at multiple PN orders~\cite{BS93,B98tail,B98quad,K95,KWWi93}. Hence, from a theoretical perspective, it is natural to expect that multiple PN coefficients may get modified if the binary components interact differently in a modified theory of gravity~\cite{AlexanderYunes07a,AlexanderYunes07b,CSreview09}. Therefore, simultaneous estimation of {\it all the PN deformation parameters}, along with the GR parameters (like component mass and spins, and other extrinsic parameters such as the distance to the source) constitute a more robust test of GR, when compared to tests that deform one PN coefficient at a time.
(See, however, Ref.~\cite{Perkins:2022fhr} for a recent study on the robustness of the single-parameter tests.) The resultant parametrized tests, where more than one PN deformation parameter is measured along with GR parameters, are often called {\it multiparameter tests of GR}~\cite{AIQS06a,Gupta:2020lxa,Datta:2020vcj,Carl:multiparam}. 

Multiparameter tests of GR are less sensitive because of the high correlations among the PN deformation parameters and between them and the intrinsic GR parameters~\cite{AIQS06a,LiEtal2011,TOG-GW150914}. Therefore, single-parameter tests are performed where one of the PN deformation parameters is measured, along with the GR parameters, fixing the other deformation parameters to their GR value, zero~\cite{AIQS06b,LiEtal2011,TOG-GW150914,GWTC-TGR,Abbott:2020jks}. 

It has been shown that the brightest prospect for carrying out multiparameter tests is by multiband observations of binary black holes~\cite{Gupta:2020lxa,Datta:2020vcj}. Multibanding involves combining signals from the same compact binary observed in the low frequency band of the space-based GW detector LISA~\cite{LISA2017} and high frequency bands of next-generation ground-based detectors like Einstein Telescope (ET)~\cite{Punturo:2010zz} or Cosmic Explorer (CE)~\cite{2019arXiv190704833R}. However, performing the multiparameter tests with a single next-generation detector remains challenging because of the aforementioned reasons.

One way to address this problem is to identify a new set of basis functions in terms of which the variance-covariance matrix of the problem is diagonal. We can do this by using the method of Principal Component Analysis (PCA) on the variance-covariance matrix corresponding to the PN deformation parameters after marginalizing over the GR parameters~\cite{AP12, methods-PCAO1O2}. 
The eigenvectors corresponding to the smallest eigenvalues provide the best measured (smallest error bars) parameters, which are linear combinations of the original PN deformation parameters. 

In this paper, we diagonalize the inverse of the covariance matrix, called the Fisher information matrix, to identify the principal components. In the diagonalized Fisher matrix, the most informative eigenvectors have the highest eigenvalues. Hence, ranking the eigenvalues in decreasing order and identifying the corresponding eigenvectors help us deduce the set of most informative new deformation parameters. The newly constructed parameters are linear combinations of the original ones and are zero in GR. We will call these new parameters as PCA parameters in the rest of the paper.

The trace of the diagonalized Fisher matrix (sum of the eigenvalues associated with the PCA parameters) captures the total amount of information it carries. The ratio of the eigenvalues to the trace of the Fisher matrix helps truncate the matrix to a specified accuracy. In summary, PCA provides a way to optimize the multiparameter tests and extract maximum information from the data with fewer additional ‘null’ deformation parameters.

In a recent paper~\cite{methods-PCAO1O2}, we demonstrated the use of this method to carry out multiparameter tests on GW events detected during the first (O1) and second observing (O2) runs of Advanced LIGO and Advanced Virgo, within the framework of Bayesian inference. We also discussed the effectiveness of the PCA parameters in detecting GR violations with simulated non-GR signals. The GR violations were introduced at every PN order starting from 1.5PN to 3.5PN. We found that the most dominant PCA parameter could recover the injected non-GR values by excluding the GR value (zero) with high credibility. However, due to the limited sensitivity of Adv LIGO and Virgo during O1/O2, we could only construct the PCA parameters by simultaneously varying deformations from 1.5PN to 3.5 PN (six-parameter tests).  Recently,  an independent work~\cite{HanSVD} also explored this method in the context of binary neutron star merger GW170817~\cite{GW170817} albeit with a different parametrization and procedure.

In this paper, we investigate the application of PCA-based multiparameter tests with the proposed next-generation ground-based detectors, Cosmic Explorer (CE)~\cite{2019arXiv190704833R} and Einstein Telescope (ET)~\cite{ETScience11} to be built in the United States and Europe, respectively. The triangular shaped ET has a noise power spectral density similar to the CE, with a better sensitivity at frequencies less than 5 Hz. The improved sensitivity of these detectors should allow us to perform a powerful eight-parameter test and compute the projected bounds on the corresponding dominant PCA parameters, as opposed to the six-parameter test available to the current generation of detectors. We use the IMRPhenomD waveform model for this study which comprises inspiral, merger, and ringdown phases of the binary evolution and considers spinning but non-precessing configurations~\cite{Khan2016}.

More specifically, we ask and answer the following questions regarding the PCA-assisted tests of GR using CE and ET:
\begin{enumerate}
    \item How many of the PCA parameters can be measured with better than 10\%  accuracy, and how does this number depend on the total mass of the binary?
    \item How does the performance of PCA-based multiparameter tests compare for the eight-parameter and six-parameter cases?
    \item What is the relative contribution of each of the original PN deformation parameters to the PCA parameters, and how do they vary with the system's total mass?
    \item How can one decide the number of PCA parameters required to capture the information in the signal up to a given accuracy, and how does this number vary with the total mass of the binary system?
\end{enumerate}

The rest of the paper is organized as follows. In Sec.~\ref{sec:parametrizedtest}, we elaborate on parametrized tests of GR and the details related to the parameter space of the signal. In Sec.~\ref{sec:FIM}, we describe our method of performing PCA on the Fisher information matrix, constructing the linear combinations of the original PN deformation parameters to build new ones, and computation of bounds on the new PCA parameters. In Sec.~\ref{sec:bounds} we study the bounds obtained on the PCA parameters as a function of total mass and investigate the sensitivity of these new linear combinations on PN deformation parameters. Finally, we investigate the hierarchy in the significance of these new PCA parameters in Sec.~\ref{sec:hier} followed by a summary of results in Sec.~\ref{sec:conclusion}.
%%%%%%%%%%%%%%%%%%%%%%%%%%%%%%%%%%%%%%%%%%%%%%%%%%%%%%%%%%%%%%%%%%%%%%%%%%%%%%%%%
%%%%%%%%%%%%%%%%%%%%%%%%%%%%%%%%%%%%%%%%%%%%%%%%%%%%%%%%%%%%%%%%%%%%%%%%%%%%%%%%%
\section{Parametrized tests of GR with inspiral-merger-ringdown waveforms}
\label{sec:parametrizedtest}
The compact binary systems emit gravitational waves as they inspiral towards each other and finally coalesce. In the frequency domain, the gravitational waveform can be schematically written as
\begin{equation}\label{eq:GWwaveform}
{\tilde h}(f)={\cal A}(f) \, e^{i\Phi (f)},
\end{equation}
where, $\mathcal{A}(f)$ and $\Phi(f)$ are the amplitude and phase of the gravitational waves. The amplitude $\mathcal{A}(f)$ depends on the source's intrinsic parameters such as component masses ($m_1$ and $m_2$), dimensionless spins ($\chi_1$ and $\chi_2$) and the extrinsic parameters like the luminosity distance ($D_L$), location and orientation of the source binary with respect to the interferometer. In the leading order, the amplitude of the gravitational waveform is proportional to $\mathcal{M}^{5/6}D_L^{-1}f^{-7/6}$, where, $\mathcal{M} = M \,\eta^{3/5}$ is called the chirpmass, $\eta = m_1 m_2/(M)^2$ is the symmetric mass ratio and $M = m_1 + m_2$ is the total detector frame mass, or redshifted mass of the binary. The redshifted total mass is related to the source frame total mass, $M_{\rm source}$ by $M_z=M_{\rm source}(1+z)$. 

When the compact binary system components slowly inspiral towards each other, the rate of change of the orbital period is much smaller than the orbital period itself $(\dot{\omega}/\omega^2 \ll 1)$. In this regime, the phase and amplitude in Eq.~(\ref{eq:GWwaveform}) can be expressed in terms of the PN expansion parameter, $\mathrm{v}\equiv (\pi\, M\,f)^{1/3}$. The frequency domain phase evolution for the inspiral phase, constructed using inputs from the PN theory, reads as~\cite{CF94}
\begin{equation}\label{eq:schematic-PN-phasing}
\Phi(f)=2\pi f\,t_c-\phi_c+\frac{3}{128\,\eta\,\mathrm{v}^5} \sum_{k=0}^N\left[
\big( \phi_k\,+\,\phi_{\rm k\,l}\,\ln \mathrm{v} \big) \,\mathrm{v}^k \right],
\end{equation}
where, $t_c$ and $\phi_c$ are coalescence time and coalescence phase of the GW signal. The PN coefficients, $\phi_k$(k = 0, 2, 3, 4, 6, 7) and $\phi_{kl}$ (k = 5, 6), are unique functions of the component masses and dimensionless spins. The phasing formula Eq.~(\ref{eq:schematic-PN-phasing}) in GR has been computed~\cite{Bliving2014} up to 3.5PN, with BH spins aligned with the orbital angular momentum of the binary.

A modified theory of gravity can differ from GR in several ways. These include the presence of extra fields, higher dimensions, and violation of diffeomorphism invariance~\cite{Berti:2015itd,YYP2016}.
Modifications due to these may show up in the conservative and the dissipative dynamics of the system leading to modifications to the GR phase evolution Eq.~(\ref{eq:schematic-PN-phasing}), leading to a modified set of PN coefficients, $\phi_k$ and $\phi_{kl}$. One of the most generic ways to test for these deviations is using a parametrized waveform model, which has deformation parameters at every PN order. The PN coefficients under this null-parametrization can be written as,
\begin{subequations}\label{eq:dev}
\begin{eqnarray} 
\phi_k  \rightarrow  \phi_k^{\rm GR}( 1+\delta\hat{\phi}_k ),\\
\phi_{kl}  \rightarrow  \phi_{kl}^{\rm GR}( 1+\delta\hat{\phi}_{kl} ).
\end{eqnarray}
\end{subequations}
Here, $\delta\hat{\phi}_k$ and $\delta\hat{\phi}_{kl}$ are the fractional non-GR deformation parameters. We do not consider deformation to the non-logarithmic part of the 2.5PN coefficient as it is not frequency dependent and hence can be absorbed into the redefinition of $\phi_c$~\cite{B95,AISS05}. Hence, we have eight non-GR PN deformation parameters to constrain from the signal in the data. These PN deformation parameters encapsulate the deviations from GR and hence take zero value in GR~\cite{LiEtal2011, Agathos15}. If these parameters measured by matching the above-mentioned parameterized waveforms with the GW data are found to be consistent with zero, then we may say the dynamics of the corresponding compact binary merger is consistent with the predictions of GR.

We employ IMRPhenomD~\cite{Khan2016} waveform for this study. These semi-analytical waveforms account for the inspiral, merger, and ringdown phases of the evolution of the binary in GR. The amplitude of this family of waveform has only the leading quadrupolar mode and assumes that the binary is composed of BHs whose spins are aligned with the orbital angular momentum and hence are non-precessing. The inspiral part of the IMRPhenomD waveform follows from Eq.~(\ref{eq:schematic-PN-phasing}) correct up to 3.5PN order. Since our method involves testing only the structure of the inspiral phase as predicted by the PN approximation to GR, we only need to change the ansatz to the inspiral part of the IMRPhenomD model by introducing PN deformation parameters as shown in Eq.~(\ref{eq:dev}), while leaving the post-merger phase unchanged.

Having introduced the `null-parametrization', our parameter space for this analysis consists of seven GR and eight non-GR deformation parameters.  
\begin{equation}\label{eq:params}
    \bm{\theta}^i = \biggl\{ \ln\,D_L,\,t_c,\,\phi_c,\, \ln\,\mathcal{M}_c,\,\eta,\,\chi_1,\,\chi_2, \{\delta\hat{\phi}_k\},\,\{\delta\hat{\phi}_{kl}\} \biggr\}.
\end{equation}
The seven GR parameters include the source's intrinsic and extrinsic parameters. The curly brackets enclosing $\delta\hat{\phi}_k$ and $\delta\hat{\phi}_{kl}$ denote the set of deformation parameters corresponding to the non-log and log types.
%%%%%%%%%%%%%%%%%%%%%%%%%%%%%%%%%%%%%%%%%%%%%%%%%%%%%%%%%%%%%%%%%%%%%%%%%%%%%%%%%
%%%%%%%%%%%%%%%%%%%%%%%%%%%%%%%%%%%%%%%%%%%%%%%%%%%%%%%%%%%%%%%%%%%%%%%%%%%%%%%%%
\section{Fisher information matrix and the Principal component analysis.}
\label{sec:FIM}
Fisher information matrix formalism~\cite{Cramer46,Rao45} has been widely used in the GW research to forecast measurement uncertainties in the parameter estimation problems. The formalism assumes a waveform model for the astrophysical signal in question and the detector noise's power spectral density (PSD)~\cite{CF94,PW95,AISS05}. Fisher matrix is the noise-weighted inner product of the derivatives of the gravitational waveforms with respect to the physical parameters of the waveform. More explicitly, the components of the Fisher matrix can be written as~\cite{CF94,PW95,AISS05,Vallisneri07},
\begin{equation}\label{eq:noise-weighted}
\Gamma_{mn}=\Big \langle\frac{\partial {\tilde h}(f)}{\partial
\theta^m},\frac{\partial {\tilde h}(f)}{\partial \theta^n} \Big \rangle
\end{equation}
where, the noise weighted inner product is defined as,
\begin{equation}
    \langle a| b\rangle =2 \,\int_{f_{\rm low}}^{f_{\rm \rm high}}\frac{a(f)\,b(f)^{*}+a(f)^{*}\,b(f)}{S_n(f)}\,df.\label{eq:int}
\end{equation}
We consider the noise PSD, $S_n(f)$ for CE given in~\cite{Kastha2018} and choose the lower frequency cutoff at 5 Hz. The noise PSD for ET is taken from~\cite{ET-D}, and the lower frequency cutoff is set at 1 Hz. The upper frequency cutoffs for both the detectors are chosen such that the ratio between the characteristic amplitude $(2\sqrt{f}|\tilde{h}(f)|)$ and the CE/ET noise amplitude spectral density is at most 10\% \cite{Datta:2020vcj, Gupta:2020lxa}.

In the limit of a high signal-to-noise ratio and stationary and Gaussian detector noise, the error bars on various parameters are given by the square root of the diagonal elements of the inverse of the Fisher matrix (called the variance-covariance matrix). This is expected to be a reasonable approximation to the errors one would get from numerically sampling the likelihood function~\cite{Balasubramanian:1995ff}. (See \cite{Vallisneri07} for a detailed discussion on possible caveats.)

In our problem, the dimensionality of the parameter space is 15: seven GR parameters ($\vec{\theta}_{\mathrm{GR}}$) and eight deformation parameters $\vec{\theta}_{\mathrm{NGR}}$. The parameters of our primary interest here are the set of eight PN deformation parameters, $\vec{\theta}_{NGR} = \{\delta\hat{\phi}_0, \delta\hat{\phi}_2, \delta\hat{\phi}_3, \delta\hat{\phi}_4, \delta\hat{\phi}_{5l}, \delta\hat{\phi}_6, \delta\hat{\phi}_{6l}, \delta\hat{\phi}_7 \}$. The Fisher matrix corresponding to this eight dimensional space can be obtained by marginalizing the full 15 dimensional Fisher matrix in Eq.~(\ref{eq:noise-weighted}) over the seven GR parameters,  $\vec{\theta}_{GR}$. We use the Schur complement method to marginalize out the GR parameters~\cite{Datta:2020vcj,Gupta:2020lxa}. The marginalized p-dimensional Fisher matrix, $\tilde{\Gamma}$ can be calculated as,
\begin{equation}\label{eq:schur}
    \tilde{\Gamma}_{p\times p} = \Gamma_{p\times p} - \Gamma_{p\times q} \Gamma^{-1}_{q\times q} \left(\Gamma^{\rm T}\right)_{q\times p},
\end{equation}
where, $\Gamma_{p\times p}$ and $\Gamma_{q\times q}$ are the non-GR and GR parameter sub-matrices of the full Fisher matrix, respectively, and $\Gamma_{p\times q}$ denotes cross-terms between the two blocks. 

The variance-covariance matrix for the non-GR parameters is the inverse of this eight dimensional marginalized Fisher matrix, $\tilde{\Gamma}$. It represents an eight dimensional error ellipsoid in a space spanned by the deformation parameters $\vec{\theta}_{NGR}$. The correlations make the Fisher matrix ill-conditioned and hence non-invertible, leading to difficulties in carrying out the multiparameter tests.

Principal component analysis offers a way of addressing the problem of correlations. We compute the eigenvalues of the Fisher information matrix and keep only the dominant few that carry most of the information. The corresponding eigenvectors provide the new deformation parameters, which are the best-measured linear combination of the original ones. The number of new parameters retained will depend on the accuracy with which we want to represent the given data. These new parameters are uncorrelated by construction and lead to a diagonal Fisher matrix in the new representation. 

We re-write the $8\times8$ matrix $\tilde \Gamma$ in terms of basis vectors that diagonalize $\tilde \Gamma$, which reads
\begin{equation}\label{eq:PCA-decomposition}
\tilde\Gamma = \mathrm{U}\,\mathrm{S}\,\mathrm{U^{T}},
\end{equation}
where, $\mathrm{U}$ is the orthogonal transformation matrix whose columns are the eigenvectors of $\tilde \Gamma$ and $\mathrm{S}$ is the diagonal singular matrix whose elements are the eigenvalues. The relative magnitude of the eigenvalues helps us determine the dominant linear combinations of the deformation parameters, which we will call PCA parameters in this paper. Equation (\ref{eq:PCA-decomposition}) represents the transformation of the highly correlated Fisher matrix,  $\tilde\Gamma$, to a space where it is diagonal (represented by $\mathrm{S}$) with the basis that are linear combinations of the original deformation parameters:
\begin{equation}\label{eq:linearcomb}
\delta\hat{\phi}^{(i)}_{\mathrm{PCA}} = \sum_k \mathrm{U}^{ik} \delta\hat{\phi}_k,
\end{equation}
where, $\mathrm{U}^{ik}$ are the elements of the $i$th column of the transformation matrix U, corresponding to the $i$th dominant PCA parameter. They determine the  relative weights of the original PN deformation parameters in the PCA parameters.
%where, $\mathrm{U}^{ik}$ are the elements of the transformation matrix U and they determine the  relative weights of the original PN deformation parameters in the PCA parameters.
The values of the coefficients $\mathrm{U}^{ik}$ depend on the properties of the source binary and is studied in detail in  Sec.~\ref{sec:bounds}. 
 
As the Fisher matrix is now in the diagonal form, the corresponding covariance matrix can be easily computed. The 1-$\sigma$ errors on the PCA parameters are given by,
\begin{equation}
\Delta[\delta\hat{\phi}^{(i)}_{\mathrm{PCA}}] =  \sqrt{\frac{1}{\mathrm{S}^{ii}}}.
\end{equation}
As mentioned earlier, the diagonal matrix $\mathrm{S}$ may be truncated depending on the relative dominance of the eigenvalues and the accuracy requirement of the problem. The ratio of the eigenvalues is also expected to depend on the source properties, such as masses and spins, and are discussed in detail in Sec.~\ref{sec:hier}. We propose retaining as many of the PCA parameters as is required to capture at least 99\% of the information in the marginalized non-GR Fisher matrix, $\tilde{\Gamma}$. This procedure ensures that the eigenvectors which carry the maximum information are kept and discard those that make the Fisher matrix ill-conditioned. 

In the following section, we will employ the method described above to compute the bounds on the new PCA parameters and investigate how our ability to measure them with CE varies with the total mass of the source compact binary.
%%%%%%%%%%%%%%%%%%%%%%%%%%%%%%%%%%%%%%%%%%%%%%%%%%%%%%%%%%%%%%%%%%%%%%%%%%%%%%%%%
%%%%%%%%%%%%%%%%%%%%%%%%%%%%%%%%%%%%%%%%%%%%%%%%%%%%%%%%%%%%%%%%%%%%%%%%%%%%%%%%%
\section{Projected bounds on the new PCA parameters %with Cosmic Explorer and Einstein Telescope
}\label{sec:bounds}
%%%%%%%%%%%%%%%%%%%%%%%%%%%%%%%%%%%%%%%%%%%%%%%%%%%%%%%%%%%%%%%%%%%%%%%%%%%%%%%%%
In this section, we use the formalism explained in Sec.~\ref{sec:FIM} to investigate the variation in the bounds of the PCA parameters as a function of the total mass of the binary for CE and ET sensitivities. We also study the composition of the PCA parameters in terms of the original PN deformation parameters.
%%%%%%%%%%%%%%%%%%%%%%%%%%%%%%%%%%%%%%%%%%%%%%%%%%%%%%%%%%%%%%%%%%%%%%%%%%%%%%%%%
\subsection{Variation of the PCA parameters with total mass}
\label{sec:bounds_mass}
\begin{figure}[htp]
\centering
\includegraphics[width=0.45\textwidth]{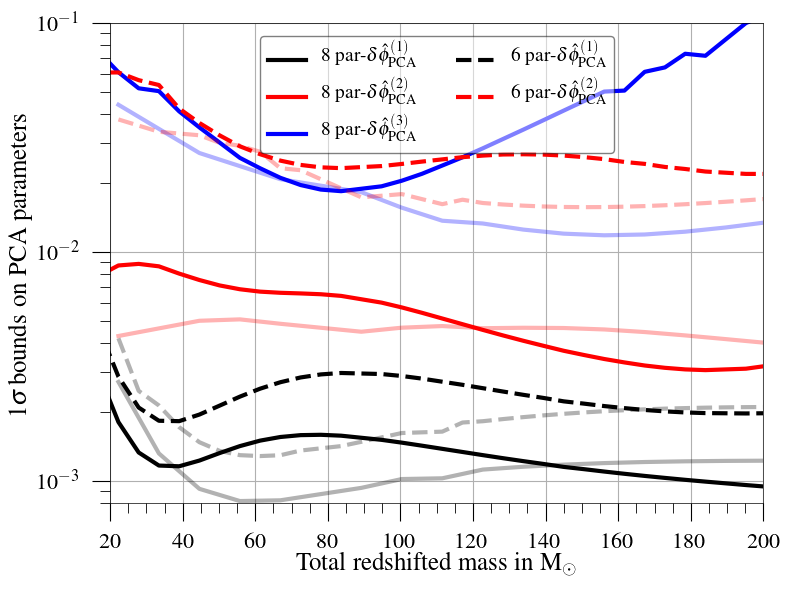}
\caption{The solid (fainter solid) lines show errors on the three most dominant PCA parameters for the 8-parameter case and the dashed (fainter dashed) lines show the errors on the two most dominant PCA parameters from the 6-parameter case, computed with CE (ET), as a function of total redshifted mass $M_z$.}
\label{fig:errors_PCA}
\end{figure}
We assume all binaries to have a mass ratio $q$ equal to 2 and individual dimensionless spins, $\chi_1$ and $\chi_2$ set to 0.2 and 0.1, respectively. We consider sources at a luminosity distance of 500 Mpc, with their redshifted total masses $M_z$ between 20--200$\,M_{\odot}$. Binary systems in this mass range have the highest SNR  $\mathcal{O} (10^2)$--$\mathcal{O} (10^3)$ accumulated during the inspiral phase and hence are ideal for this inspiral-based tests of GR.

We consider two kinds of multiparameter tests: (1) 8-parameter test, where we measure all the eight PN deformation parameters simultaneously, and (2) 6-parameter test, where only the last six PN deformation parameters (1.5PN to 3.5PN) are measured together. Fig.~\ref{fig:errors_PCA} shows the projected 1-$\sigma$ bounds on the PCA parameters as a function of the total mass of the source compact binary. The solid (dashed) lines represent the bounds on the first three (two) dominant PCA parameters for the 8-parameter (6-parameter) test computed with CE. The fainter lines denote the same quantities computed with ET.

We will first discuss the bounds obtained with CE. We find that for the 8-parameter test, the three most dominant PCA parameters are measurable with 1-$\sigma$ errors less than 10\%. For the 6-parameter test, we can constrain only two of the dominant PCA parameters with a similar error bar. The most dominant PCA parameter, $\delta\hat{\phi}^{(1)}_{\mathrm{PCA}}$ for both the 8-parameter and 6-parameter cases have the best bounds of $\mathcal{O}(10^{-3})$.  Hence, we will call $\delta\hat{\phi}^{(1)}_{\mathrm{PCA}}$  the best measured linear combination for the corresponding multiparameter test.
We do not show the bounds on the rest of the sub-dominant PCA parameters, as they mostly capture noise features and are uninformative.

The errors on the dominant PCA parameter for both the 8-parameter (solid black line) and 6-parameter (dashed black line) cases have similar features, as shown in Fig.~\ref{fig:errors_PCA}. The bounds slowly fall off with an increasing total mass of the system with  a minimum around $\sim 35\,M_{\odot}$. The gradual improvement in the constraints occurs for the following reasons: (1) Increase in the inspiral SNR with the increasing total mass of the system. (2) The late time dynamics is more prominent for higher mass systems, making the higher PN orders carry significant information. Both these effects help in breaking correlations between the lower order PN deformation parameters with chirp mass and mass ratio (largely between $M_c - \delta\hat{\phi}_0 $, $\eta - \delta\hat{\phi}_0 $, $\eta - \delta\hat{\phi}_2 $  and $\eta - \delta\hat{\phi}_3 $). In the next subsection, we explore the varying degree to which this happens across the mass range and influences the bounds on the PCA parameters. The minimum around $\sim 35\,M_{\odot}$ results from a complex interplay between the two factors mentioned above and the worsening of the errors above $\sim 35\,M_{\odot}$ is caused by the decrease in the number of inspiral cycles with total mass.

The above two reasons also apply to the bounds on the second-dominant PCA parameter $\delta\hat{\phi}^{(2)}_{\mathrm{PCA}}$ for both the 8-parameter (solid pink line) and 6-parameter (dashed pink line) cases, which also decreases with increasing total mass. However, the improvement in the bounds is steeper compared to the dominant PCA parameter. This feature can be attributed to the second dominant PCA parameter being more sensitive to the higher order effects (sub-dominant dynamics of the system), which only become more pronounced for high mass systems. We will study this aspect in more detail in the following subsection.

The bounds on the dominant PCA parameter for the 8-parameter case are better than the bounds from the 6-parameter case for every system. This is because in the 8-parameter case, $\delta\hat{\phi}^{(1)}_{\mathrm{PCA}}$ is a linear combination of all the eight PN deformation parameters, which includes the lowest two PN deformation parameters $\delta\hat{\phi}_0$ and $\delta\hat{\phi}_2$. As these two lowest order PN deformation parameters are well measured and mostly contribute to the most dominant PCA parameter, they yield better bounds than the six parameter case. The same argument applies for the second-dominant PCA parameter, $\delta\hat{\phi}^{(2)}_{\mathrm{PCA}}$, which is also a function of these two PN parameters and provides a stronger bound in the eight parameter case.

The order of magnitude estimates in measuring the PCA parameters with ET do not differ much from CE. For the 8-parameter (6-parameter) case, the bounds on the leading PCA parameter with ET, improve by a factor of 2--3 (1.5) for masses between 40--$100\,M_{\odot}$, as compared to CE. The bounds on the third-dominant PCA parameter for the 8-parameter case improve the most---by a factor of 2--9, for higher mass systems ($\gtrsim 100 M_{\odot}$), as shown in Fig.~\ref{fig:errors_PCA}. This improvement for ET compared to CE is due to the improved low frequency sensitivity of the former, leading to an increase in the number of cycles between 1 Hz and 5Hz.

%%%%%%%%%%%%%%%%%%%%%%%%%%%%%%%%%%%%%%%%%%%%%%%%%%%%%%%%%%%%%%%%%%%%%%%%%%%%%%%%%
\subsection{Sensitivity of the PCA parameters to the different PN deformation parameters}
\label{sec:PCAcoeffs_mass}
\begin{figure*}
\begin{minipage}[b]{1\textwidth}
\includegraphics[scale=0.6]{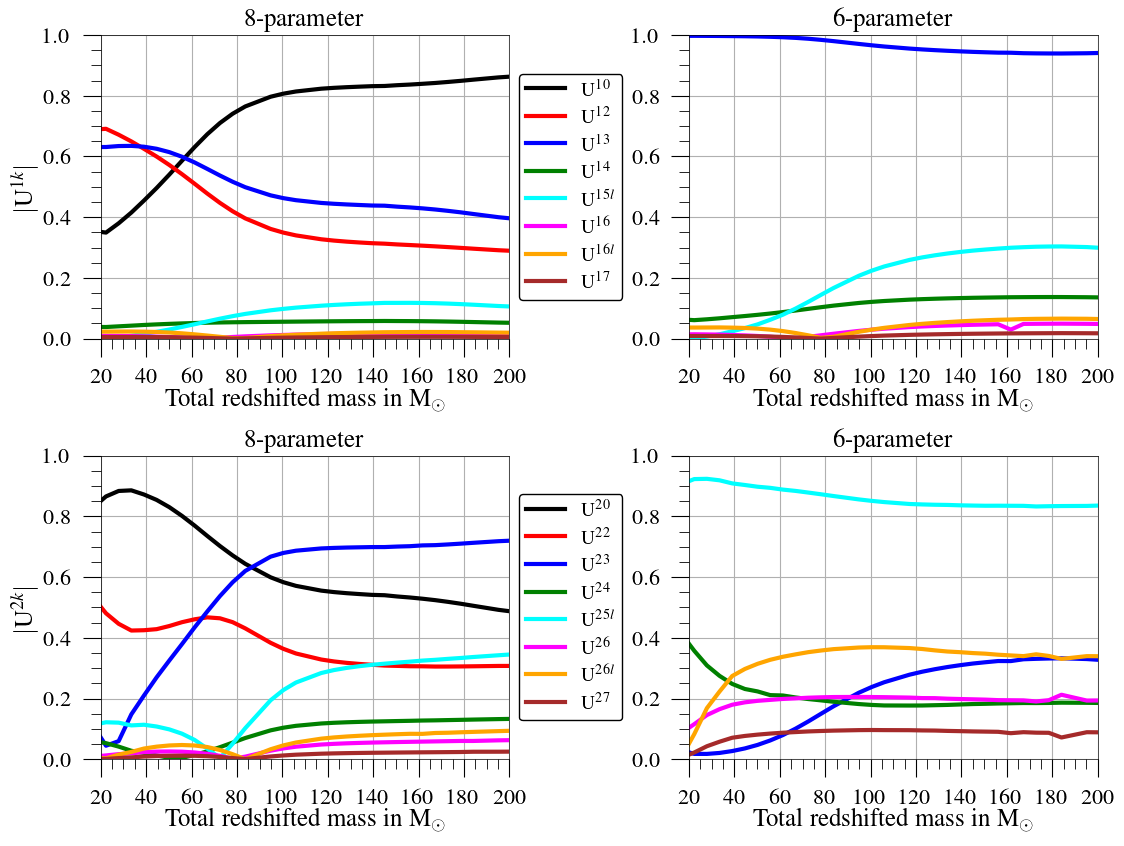}
\caption{Absolute values of the coefficients of the most dominant (top row) and second-dominant (bottom row) PCA parameter from 8-parameter (left column) and 6-parameter (right column) cases, as a function of redshifted total mass of the systems, $M_z$.}
%{\change y label and legends to $\alpha_{1k}$}} 
\label{fig:hierPCA}.
\end{minipage}
\end{figure*}
Next, we explore in depth the composition of the PCA parameters. The PCA parameters are optimal linear combinations of the PN deformation parameters, as shown in Eq.~(\ref{eq:linearcomb}). In this section, we study the relative contribution of the original PN deformation parameters to the bounds on the PCA parameters and the variation with an increasing total mass of the systems. We can study the composition of the PCA parameters by comparing the magnitude of the elements of the transformation matrix $\mathrm{U}^{ik}$ shown in Eq.~(\ref{eq:linearcomb}). 

The transformation matrix is unique for every system and strongly depends on the total mass. The structure of the PCA parameters will also tell us about the dynamics they probe. For example, a PCA parameter which is composed of coefficients with large magnitudes corresponding to lower order PN deformation parameters, like $\delta\hat{\phi}_0$, $\delta\hat{\phi}_2$ and $\delta\hat{\phi}_3$, and negligible corresponding to the higher order deformation parameters, can only probe the early inspiral phase. 

We next study the composition of the different PCA parameters in terms of the original PN ones in the case of CE. The plots in the left and right columns of the top row of Fig.~\ref{fig:hierPCA} show the variation in the magnitude of the coefficients of the dominant PCA parameter for the 8-parameter and 6-parameter cases, respectively. The top-left panel of Fig.~\ref{fig:hierPCA} clearly shows that for the 8-parameter case, only 0PN ($\delta\hat{\phi}_0$), 1PN ($\delta\hat{\phi}_2$) and 1.5PN ($\delta\hat{\phi}_3$) deformation parameters contribute significantly to the bounds on the dominant PCA parameter, $\delta\hat{\phi}^{(1)}_{\mathrm{PCA}}$. Similarly, from the top-right panel of Fig.~\ref{fig:hierPCA}, we see that the most dominant PCA parameter for the 6-parameter case is mostly sensitive to the 1.5PN ($\delta\hat{\phi}_3$) and 2.5PN-log ($\delta\hat{\phi}_{5l}$) deformation parameters.

From the values of $|\mathrm{U}^{1k}|$ in the top-left panel of Fig.~\ref{fig:hierPCA}, 1PN and 1.5PN deformation parameters dominate over 0PN for lower masses (up to $\sim 50\,M_{\odot}$). With increasing total mass, 0PN rises in dominance and dictates the overall characteristics of the bounds on $\delta\hat{\phi}^{(1)}_{\mathrm{PCA}}$. This happens because, for lower masses, the contribution from the higher PN orders is not enough to break the degeneracy between chirp mass and 0PN deformation parameter, $\delta\hat{\phi}_0$. For higher mass systems, the higher order PN coefficients become relevant to break the $M_c - \delta\hat{\phi}_0$ degeneracy along with an increase in the SNR. This causes the 0PN deformation parameter to influence the bounds on the 8-parameter-$\delta\hat{\phi}^{(1)}_{\mathrm{PCA}}$ the most (denoted by solid black line in Fig~\ref{fig:errors_PCA}).

The relative values of $|\mathrm{U}^{1k}|$ in the top-right panel of Fig.~\ref{fig:hierPCA} convey that, for the 6-parameter case, it is the 1.5PN deformation parameter that has the strongest influence on the measurability of the most dominant PCA parameter. However, as the total mass of the system increases, the 2.5PN-log deformation parameter rises in significance, although its relative contribution is almost three times lower compared to the 1.5PN deformation parameter, even for the highest masses.

The variation in the magnitude of the coefficients of the second dominant PCA parameter, $\delta\hat{\phi}^{(2)}_{\mathrm{PCA}}$ for the 8-parameter and 6-parameter cases are shown in the bottom-left and bottom-right panels of Fig.~\ref{fig:hierPCA}, respectively. First important observation from the bottom-left panel of Fig.~\ref{fig:hierPCA} is that, 0PN, 1PN, 1.5PN and 2.5PN-log deformation parameters are the dominant contributors to the 8-parameter $\delta\hat{\phi}^{(2)}_{\mathrm{PCA}}$. 
Secondly, the contribution of $\delta \hat{\phi}^{(0)}$ decreases with increasing mass in contrast to what we saw for the dominant PCA parameter, $\delta\hat{\phi}^{(1)}_{\mathrm{PCA}}$. The values of $|\mathrm{U}^{23}|$ and $|\mathrm{U}^{25l}|$ sharply increase as a function of mass, indicating increased sensitivity of  $\delta\hat{\phi}^{(2)}_{\mathrm{PCA}}$ to higher PN effects.

The bottom-right panel of Fig.~\ref{fig:hierPCA} tells that for the 6-parameter case, all the deformation parameters starting from 1.5PN contribute to the bounds on $\delta\hat{\phi}^{(2)}_{\mathrm{PCA}}$, with the exception of $|\mathrm{U}^{27}|$ which contributes the least. The bounds on the second dominant PCA parameter for the six-parameter case have the dominant contribution from the 2.5PN-log deformation parameter, unlike the dominant PCA parameter, where the 1.5PN deformation parameter contributes the most (shown in the top-right panel of Fig.~\ref{fig:hierPCA}). 

The rest of the sub-dominant PCA parameters should be sensitive to higher PN deformation parameters. However, they are uninformative because of the significant error bars associated with their measurement. This reflects the limitation of what one can do with the sensitivities of next-generation detectors. Lastly, given the similarity in the overall features between ET and CE curves, we do not repeat this for ET.
%%%%%%%%%%%%%%%%%%%%%%%%%%%%%%%%%%%%%%%%%%%%%%%%%%%%%%%%%%%%%%%%%%%%%%%%%%%%%%%%%
%%%%%%%%%%%%%%%%%%%%%%%%%%%%%%%%%%%%%%%%%%%%%%%%%%%%%%%%%%%%%%%%%%%%%%%%%%%%%%%%%
\section{Determining the number of significant PCA parameters}\label{sec:hier}

We have seen in the previous section that the principal components are unique to each binary and are determined by the specific parameters characterizing the system. Hence, it is only natural to expect that the relative dominance of the PCA parameters will also depend on the source properties.  Evaluation of the relative dominance of the PCA parameters will help us determine the number of significant PCA parameters needed for an accurate reconstruction of the information in the signal. In this section we will describe our proposal to determine the number of principal components that carry most of the information relevant to parameterized tests of GR.

The eigenvalues of the Fisher information matrix encode the relative importance of the eigenvectors in terms of the amount of information they carry. 
The trace of the diagonal Fisher matrix captures the total information contained in the signal and it is invariant under the transformation in Eq. (\ref{eq:PCA-decomposition}).
For any given system, let $\{\lambda_k\},$ $k=1,\ldots,n,$ denote the eigenvalues obtained after finding the PCA parameters from the $n$-dimensional marginalized Fisher matrix calculated from Eq.~(\ref{eq:schur}). We shall further assume that the eigenvalues have been ordered such that $\lambda_1 \ge \lambda_2  \ge \ldots \ge \lambda_n$. The relative information carried by the $k^{\rm th}$ eigenvector is calculated by dividing the corresponding eigenvalue by the trace of the matrix:
\begin{equation}\label{eq:POV}
I_k = \frac{\lambda_k}{\sum_{k=1}^{n} \lambda_k}.
\end{equation}
The cumulative information carried by the $m$ dominant PCA parameters is calculated as:
\begin{equation}\label{eq:cumPOV}
    \sum_{k=1}^{m} I_k = \frac{\sum_{k=1}^{m} \lambda_k}{\sum_{k=1}^{n} \lambda_k}\,\,\,\,\,(m\leq n).
\end{equation}
\begin{figure}[htp]
\centering
\includegraphics[scale=0.4]{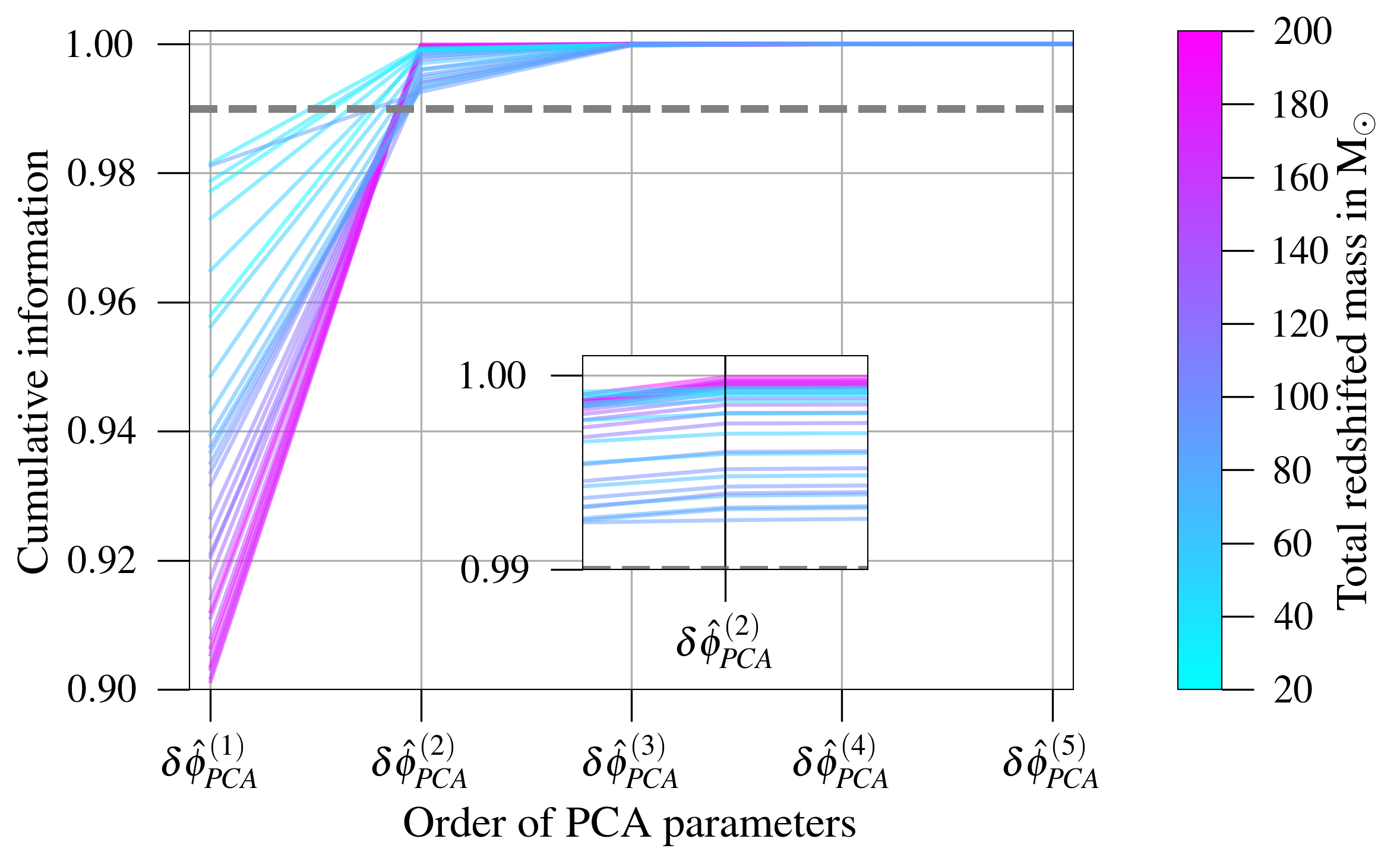}
\caption{Cumulative information carried by the PCA parameters  as defined in Eq.~\ref{eq:cumPOV} for different masses. Dashed horizontal lines indicate 99\% information.}
\label{fig:cumPOV}
\end{figure}

We calculate the cumulative information content of the PCA parameters according to Eq.~(\ref{eq:cumPOV}) for all the systems and plot them as a function of mass in Fig.~\ref{fig:cumPOV} (These plots are called Scree plots~\cite{Screeplot}). We use these plots to determine the number of significant PCA parameters for a given system depending on the amount of information we want to capture. We set a threshold of 99\% for the total information to be retained by the PCA parameters denoted by the grey shaded line.

The cumulative information monotonically increases as we include more sub-dominant PCA parameters. The sharpest increase in the information is between the most-dominant and the second-most-dominant PCA parameters. The increment beyond this point is relatively slower and quickly saturates to values very close to one as more of the sub-dominant PCA parameters are included. This is why we have shown only the five most dominant PCA parameters in the plot. 

The amount of information captured by $\delta\hat{\phi}^{(1)}_{PCA}$ gradually increases from high to low mass systems. This is because the dominant PCA parameter is mostly influenced by the first three PN deformation parameters ($\delta\hat{\phi}_0$, $\delta\hat{\phi}_2, \delta\hat{\phi}_3$ ) as shown in the top-left plot in Fig.~\ref{fig:hierPCA}. The GW signals become increasingly inspiral dominated with decreasing mass, allowing the lower PN orders to carry most of the information. On the other hand, $\delta\hat{\phi}^2_{PCA}$ carries more information for higher mass systems as compared to lower ones (see inset of Fig.~\ref{fig:cumPOV}). This is because high-mass systems probe more relativistic dynamics encoded in the higher order PN deformation parameters, to which $\delta\hat{\phi}^2_{PCA}$ is more sensitive as shown in the bottom-left panel of Fig.~\ref{fig:hierPCA}.

We find that the first PCA parameter captures over 90\% of the information for all the systems analyzed, but none cross the required threshold (all the points lie below the grey dashed line marking 99\% of the information). This suggests we must consider at least one of the sub-dominant PCA parameters. The most-dominant and the second-most-dominant PCA parameters together capture more than 99\% of the information as shown in the Fig.~\ref{fig:cumPOV} and hence meet our criteria.

\section{Conclusion}\label{sec:conclusion}
 We discussed the optimization of multiparameter tests of GR, where one simultaneously measures parameterized departures from GR in the GW phase evolution predicted by the PN theory, using the Principal Component Analysis. We studied this in the context of Cosmic Explorer and Einstein Telescope, proposed next-generation GW detectors. Principal component analysis helps identify the best-measured deformation parameters that are linear combinations of the original PN deformation parameters. These new PCA parameters not only have the least statistical errors but also are sensitive to multiple PN deformation parameters, making the test of GR more robust. 
 
 Below we list the most important findings of this paper which answer the questions we posed in the introduction:
\begin{enumerate}
    \item Three (two) of the most dominant PCA parameters can be constrained with a 1-$\sigma$ error bar of less than 0.1 for the case where all the eight (six) PN deformation parameters were varied simultaneously. The variation of the measurement errors with the total mass of the binary are explored in detail (in Sec.~\ref{sec:bounds_mass} and Fig.~\ref{fig:errors_PCA}).
    \item The PCA parameters for the 6-parameter test, where the six PN deformation parameters (1.5PN to 3.5PN) were varied, have weaker bounds compared to the case where all the eight PN deformation parameters are simultaneously measured. This difference arises because the 0PN and 1PN deformation parameters that are very well measured from the GW signal, are assumed to follow GR (fixed to their GR value of zero) in the 6-parameter test, unlike in the 8-parameter test, where, the contribution from 0PN and 1PN deformation parameters help in constraining the PCA parameters better. This is explained in more detail in Sec.~\ref{sec:bounds_mass}.
    \item The most dominant PCA parameter is sensitive to lower order PN deformation parameters, whereas the second dominant PCA parameter, captures higher order PN effects better with increasing mass. Fig.~\ref{fig:hierPCA} show these features and a detailed discussion can be found in Sec.~\ref{sec:PCAcoeffs_mass}.
    \item Lastly, we propose to use a truncation criterion based on the information carried by each of the PCA parameters. We retain only the PCA parameters, which can contribute to at least 99\% of the information carried by the full Fisher matrix. This criterion may be changed depending on the exact science one is interested in (see Fig.~\ref{fig:cumPOV} and Sec.~\ref{sec:hier} for details). 
\end{enumerate}

Besides these, we find that the relative weights of PN deformation parameters beyond the 2.5PN log-term are weak. This is true even in the case of six-parameter tests, which implies that the next-generation ground based detectors have limited ability to bound deviations at the higher PN end.

%%%%%%%%%%%%%%%%%%%%%%%%%%%%%%%%%%%%%%%%%%%%%%%%%%%%%%%%%%%%%%%%%%%%%%%%%%%%%%%%%
%%%%%%%%%%%%%%%%%%%%%%%%%%%%%%%%%%%%%%%%%%%%%%%%%%%%%%%%%%%%%%%%%%%%%%%%%%%%%%%%%
\acknowledgments
We thank Sebastian Khan for sharing his Mathematica code of the IMRPhenomD waveform model with us. We thank Anuradha Gupta and Arnab Dhani for very useful discussions. K.G.A., M. S., and S.D. acknowledge the Swarnajayanti grant DST/SJF/PSA-01/2017-18 of the Department of Science and Technology, India.  K.G.A and B.S.S. acknowledge the support by the Indo-US Science and Technology Forum through the Indo-USCentre for Gravitational-Physics and Astronomy, grant IUSSTF/JC-029/2016. K.G.A acknowledges the support of the Core Research Grant CRG/2021/004565 and MATRICS grant MTR/2020/000177 of the Science and Engineering Research Board of India. K.G.A and S.D also acknowledge support from Infosys Foundation. M.S. acknowledges the support from the National Science Foundation with grants PHY-1806630, PHY-2010970, and PHY-2110238. B.S.S. acknowledges the support of the National Science Foundation with grants PHYS-2012083 and AST-2006384. We use the following software packages for the computations in this work: \textsc{NumPy} \cite{vanderWalt:2011bqk}, \textsc{SciPy} \cite{Virtanen:2019joe}, \textsc{Matplotlib} \cite{Hunter:2007}. 

\bibliography{references.bib} 
\bibliographystyle{apsrev}

\end{document}